\documentclass[aps,prl,twocolumn,amsmath,amsfonts,amssymb, braket, groupedaddress,floatfix]{revtex4-2}

\usepackage{graphicx}
\usepackage{dcolumn}
\usepackage{bm}
\usepackage{hyperref}
\usepackage[usenames,dvipsnames]{color}
\usepackage[normalem]{ulem}
\usepackage{amsmath}
\usepackage{amssymb}
\usepackage{braket}

\begin{document}


\title{A unified equation for saturation magnetization and spin transport in weakly disordered ferromagnets}

\author{Sumanta Mukherjee$^{1,*}$}

\affiliation{$^1$ Solid State and Structural Chemistry Unit, Indian Institute of Science, Bengaluru, Karnataka 560012, India\\}

\email{sm31081985@gmail.com}

\date{\today}

\begin{abstract}
In this report, a unified description of the loss of saturation magnetization in the presence of a distribution of finite-size effects is provided for weakly disordered spin-1/2 ferromagnets. This description allows us to derive a unified form of the Bloch equation for these systems. We further extend this approach to obtain a unified expression for spin transport in such systems.
\end{abstract}

\maketitle

While the Mermin–Wagner theorem\cite{ref-1,ref-2} states that an infinite 1d or 2d Heisenberg ferromagnetic system cannot exhibit long-range order at any finite temperature due to the high density of low-energy excitations—whose associated entropy suppresses ordering—a 3d Heisenberg system, being above the lower critical dimension, experiences reduced fluctuations and can therefore exhibit long-range order at finite temperature\cite{ref-3}. A vast number of studies\cite{ref-4,ref-5,ref-6,ref-7,ref-8,ref-9,ref-10,ref-11,ref-12} have been carried out on 1d and 2d Heisenberg systems to elucidate their magnetic behavior. In this context, it has been suggested that in finite Heisenberg systems, finite-size effects can open gaps in the low-energy magnon excitation spectrum\cite{ref-13,ref-14}, resulting in an exponential suppression of the density of states for low-energy magnons. Such Lifshitz tail-like suppression\cite{ref-15,ref-16,ref-17,ref-18} of the low-energy density of states may significantly alter the magnetization behavior of 1d or 2d Heisenberg systems, often leading to ordering behavior similar to that of a 3d ferromagnetic system\cite{ref-19,ref-20}. While any real system is finite, the presence of intrinsic disorder, such as vacancies, can further fragment the lattice, thereby enhancing finite-size effects\cite{ref-3}. Since such disorder is random, its introduction into an otherwise infinite lattice results in a distribution of segment lengths\cite{ref-3}. This distribution of lengths, in turn, leads to a corresponding distribution of energy gaps in the system\cite{ref-31,ref-21}.\\
In this report, we present a comprehensive study that incorporates the distribution of energy gaps in conjunction with the magnon spectrum and describes the resulting magnetization behavior in such systems. We calculate the reduction in saturation magnetization due to magnon excitations in the presence of a gap distribution. Furthermore, we extend this analysis to different effective dimensionalities, deriving a unified equation that generalizes the well-known Bloch equation\cite{ref-22,ref-23} for magnetization loss and captures the behavior of weakly disordered spin-1/2 Heisenberg systems. We further extend the analysis to understand spin transport phenomena in such disordered systems, deriving a unified equation for spin current. We find that the final expression for the spin current in weakly disordered magnetic systems is structurally analogous to the Efros–Shklovskii law\cite{ref-24} for electronic conductivity in disordered systems. However, similar to Bloch's equation\cite{ref-23}, the presented model for magnetization loss and spin transport is valid only in the low-temperature regime, where additional complexities are minimized.\\

In a disordered system, the probability that a fragmented region of length $L$ is free of disorder and can be treated as a clean region is given by\cite{ref-3}.
\begin{equation}
P(L) = a \exp \left\{ - a L^{\psi}  \right\}, \text{ where } a = - \ln (1-p)
\end{equation}
Here, $p$ denotes the dilution or strength of the disorder, whereas $\psi$ represents the effective dimensionality\cite{ref-3}. The prefactor $a$ outside the exponential may not be required; however, it is often introduced as a normalization constant, for example for $\psi = 1$. Omitting or retaining this constant outside the exponential does not affect the outcome of the derivation, as the calculation relies on numerical normalization. On the other hand, the presence of finite size effectively discretizes the magnon spectrum in any dimension, introducing gaps $(\epsilon)$ that depend on the system length $L$ as\cite{ref-13,ref-14}.
\begin{equation}
\epsilon = \frac {c n^2}{L^2}
\end{equation}
Here, $c$ is a constant, and $n$ is an integer, typically taking values 1,2,3,$\ldots$ Note that, since there is no particular restriction on the value of $L$ within these expressions, there is consequently no restriction on the value of $\epsilon$ either. In other words, within this model, there is no critical ordering temperature of the system, leading to a smeared transition over the entire temperature range. This behavior is in complete contrast to what one would expect from Griffiths-phase-like behavior in a disordered system\cite{ref-3,ref-21}. Further, in the presence of a gap distribution arising from the distribution of lengths, one may combine these two equations to obtain the expression for the gap distribution\cite{ref-3,ref-21} at fixed $n$:
\begin{figure}[t]
\begin{center}
\includegraphics[width=1.0\columnwidth]{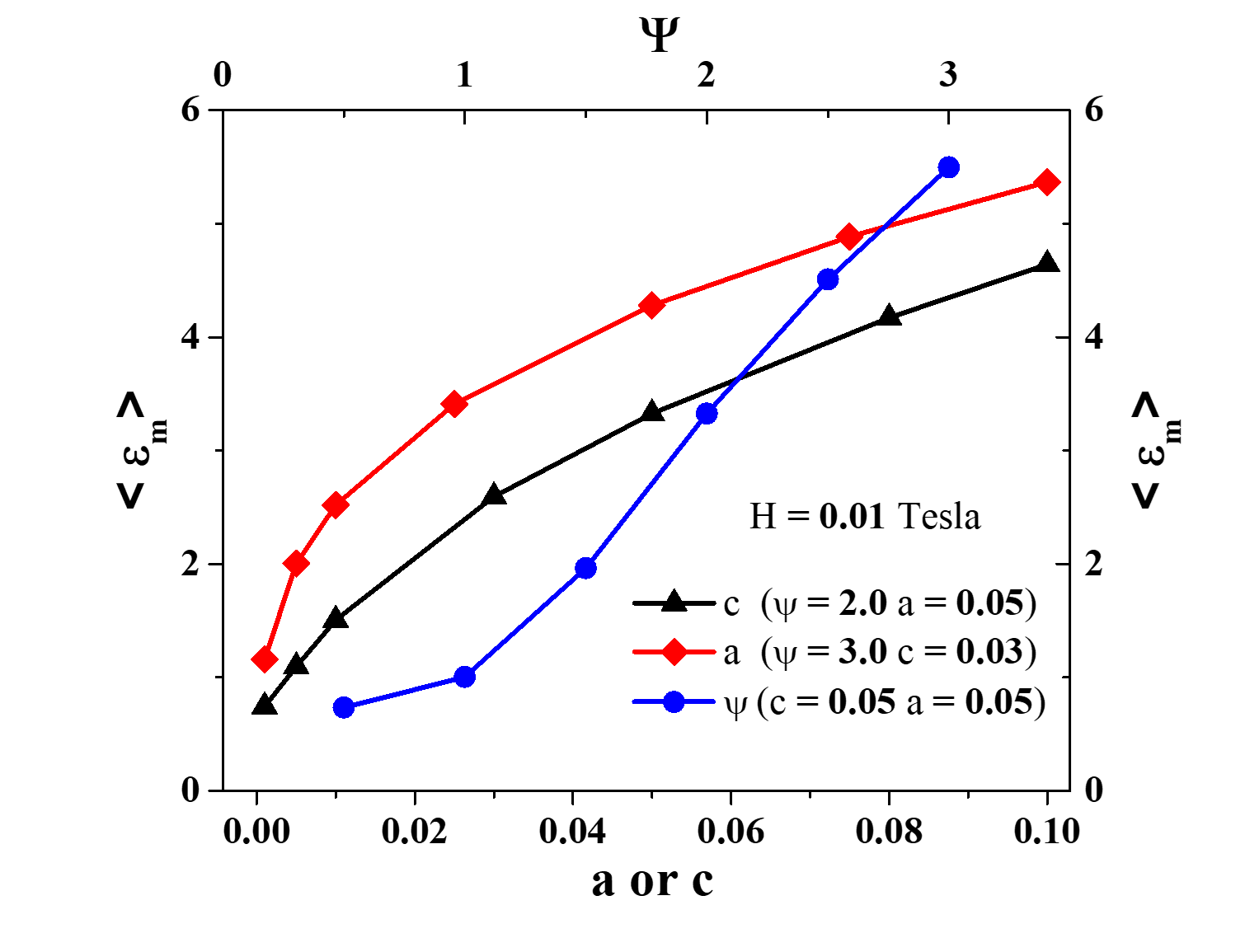}
\caption{The variation of $<\epsilon_m>$ for different values of the parameters $a$, $c$, and $\psi$ is shown.}
\label{fig1}
\end{center}
\end{figure}
\begin{equation}
P(\epsilon) = \frac {a n \sqrt{c}}{2\epsilon^{3/2}} \exp \left\{ - a \left( n \sqrt{\frac{c}{\epsilon}}\right)^{\psi} \right\}
\end{equation}
Considering different values of $n$, one may write the total probability as
\begin{equation}
P_{t}(\epsilon) = \sum_{n=1}^{n} \frac {a n \sqrt{c}}{2\epsilon^{3/2}} \exp \left\{ - a \left( n \sqrt{\frac{c}{\epsilon}}\right)^{\psi} \right\}
\end{equation}
This expression is very similar to the Lifshitz tail-like exponential suppression of the density of states, as is generally observed in disordered systems\cite{ref-15,ref-16,ref-17,ref-18}. With this, one may define the mean of this gap distribution as
\begin{equation}
\resizebox{.97\hsize}{!}
{$<\epsilon_m> = \frac{\int \epsilon P_{t}(\epsilon) d\epsilon}{\int P_{t}(\epsilon) d\epsilon} = \frac {\sum_{n=1}^{n} \int \epsilon \frac {a n \sqrt{c}}{2\epsilon^{3/2}} \exp \left\{ - a \left( n \sqrt{\frac{c}{\epsilon}}\right)^{\psi} \right\} d\epsilon}{\sum_{n=1}^{n} \int \frac { a n \sqrt{c}}{2\epsilon^{3/2}} \exp \left\{ - a \left( n \sqrt{\frac{c}{\epsilon}}\right)^{\psi} \right\} d\epsilon}$}
\end{equation}
Furthermore, in most real systems, the gap typically opens in the low-energy region of the magnon dispersion\cite{ref-13,ref-25}, while the high-energy part of the spectrum, though discrete, may be treated as effectively continuous\cite{ref-25}. Hence, the above expression can be written as
\begin{equation}
<\epsilon_m> =  \frac {\int_{1}^{n} \int_{n^2x}^{n^2y} \epsilon \frac {g a n \sqrt{c}}{2\epsilon^{3/2}} \exp \left\{ - a \left( n \sqrt{\frac{c}{\epsilon}}\right)^{\psi} \right\} d\epsilon dn}{\int_{1}^{n} \int_{n^2x}^{n^2y} \frac {g a n \sqrt{c}}{2\epsilon^{3/2}} \exp \left\{ - a \left( n \sqrt{\frac{c}{\epsilon}}\right)^{\psi} \right\} d\epsilon dn}
\end{equation}
While, in a discretized magnon spectrum, it may not be appropriate to include the usual dimension-dependent power-law density of states factor, one may instead introduce a degeneracy factor $g$ in the above expression. In a crude approximation, $g$ is at least equal to two, accounting for the symmetry of magnon modes defined by $k$ and $-k$. In the low-gap limit, this factor may be generalized to an energy-dependent form $g(\epsilon)$; for example, $g(\epsilon) \approx \epsilon^{-1/2}$ in a 1d system\cite{ref-22}. With this expression, the parameters $a$, $c$, and $\psi$ can be treated on an equal basis, as all three enter the theory in an equivalent manner and determine the mean gap $<\epsilon_m>$ of the system. \textbf{Figure 1} shows the dependence of $<\epsilon_m>$ (in eV) on each of these parameters.
With these expressions, and for a spin-1/2 system, one may calculate the magnetization (saturation) loss per spin at any finite temperature due to magnon excitations as\cite{ref-23,ref-25}.
\begin{equation}
\begin{aligned}
&M_{loss}  =  \mu_B \frac{\int <n_{\epsilon}> P_{t}(\epsilon) d\epsilon}{\int P_{t}(\epsilon) d\epsilon}\\
& = \mu_B \frac {\int_{1}^{n} \int_{n^2x}^{n^2y} <n_{\epsilon}> \frac {g a n \sqrt{c}}{2\epsilon^{3/2}} \exp \left\{ - a \left( n \sqrt{\frac{c}{\epsilon}}\right)^{\psi} \right\} d\epsilon dn}{\int_{1}^{n} \int_{n^2x}^{n^2y} \frac {g a n \sqrt{c}}{2\epsilon^{3/2}} \exp \left\{ - a \left( n \sqrt{\frac{c}{\epsilon}}\right)^{\psi} \right\} d\epsilon dn}
\end{aligned}
\end{equation}
\begin{figure}[t]
\begin{center}
\includegraphics[width=1.0\columnwidth]{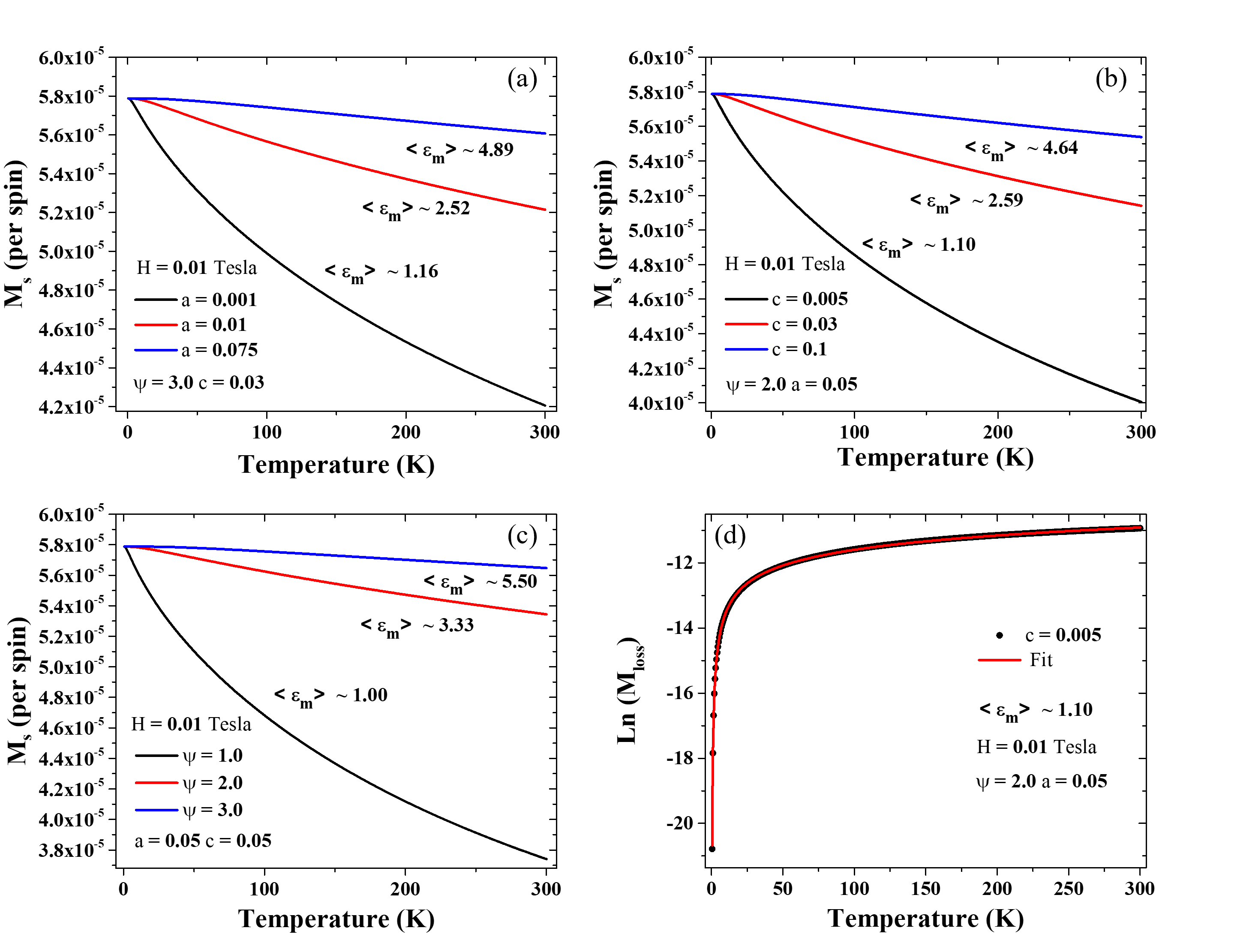}
\caption{(a)–(c) The calculated variation of the saturation magnetization ($M_s$), obtained by varying the parameters $a$, $c$, and $\psi$, respectively, at different temperatures and in a fixed low magnetic field, is shown. (d) shows representative fits of $\ln(M_{\text{loss}})$ using \textbf{Equation 11} with $F(H)=1$.}
\label{fig2}
\end{center}
\end{figure}
\begin{figure*}[t]
\begin{center}
\includegraphics[width=2.0\columnwidth]{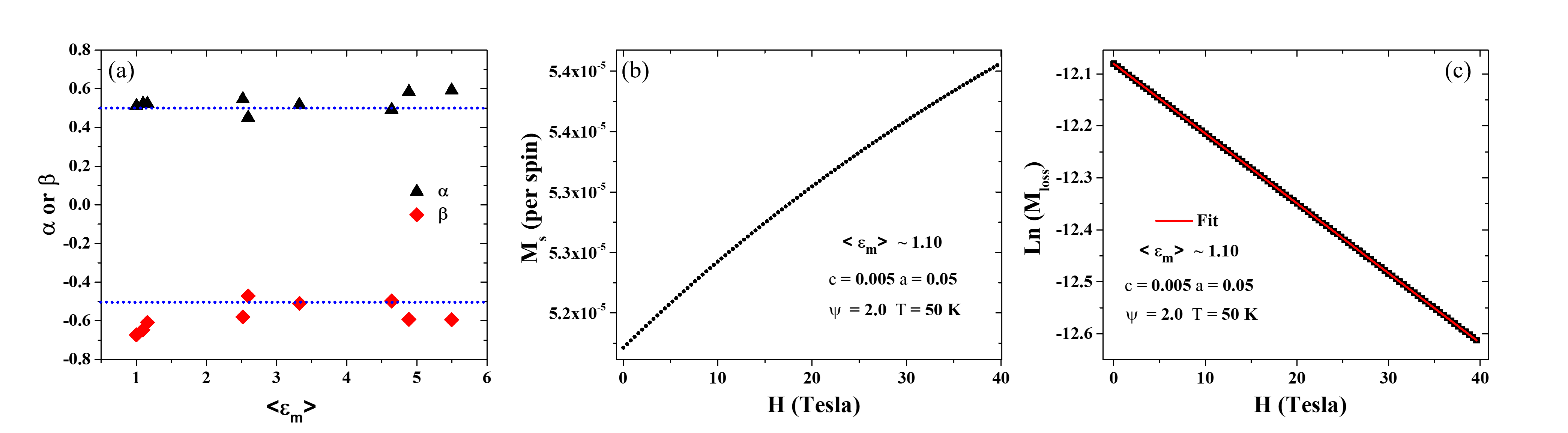}
\caption{(a) shows the typical values of the exponents $\alpha$ and $\beta$ obtained from fitting the $M_{\text{loss}}$ data using \textbf{Equation 11} for various values of $<\epsilon_m>$, obtained through controlled variation of the parameters $a$, $c$, and $\psi$. (b) A typical variation of $M_s$ as a function of magnetic field is shown. (c) A representative fit of $\ln(M_{\text{loss}}(H))$ obtained using \textbf{Equation 12} is shown.}
\label{fig3}
\end{center}
\end{figure*}
Where $\mu_B$ is the Bohr magneton, and the occupation number of a mode $<n_{\epsilon}>$ , under the discrete condition with large $<\epsilon_m>$, may be approximated by a Maxwell–Boltzmann distribution\cite{ref-22,ref-26} of the form given by
\begin{equation}
<n_{\epsilon}> = \exp\left\{-\epsilon/k_B T\right\}
\end{equation}
Where $k_B$ is the Boltzmann constant and $T$ is the temperature (note the symobol $<n_{\epsilon}>$ and $n$ are used to define different quantities).
One may also include the effect of an external magnetic field $H$\cite{ref-27}, which introduces additional modifications to the gap distribution via the approximate Zeeman shift\cite{ref-26,ref-27} within linear spin-wave theory. It can be shown that the magnetic field effectively enters through the integration limits above, as well as through the occupation factor, modifying it to become
\begin{equation}
<n_{\epsilon^\prime}> = \exp\left\{\epsilon^\prime/k_B T\right\}
\end{equation}
Where $\epsilon^\prime = \epsilon + \mu_B H$ for a spin-1/2 system\cite{ref-27,ref-28}. While one may consider nonlinear effects of the magnetic field on the magnon gap, the linear effect—often observed in experiments and theory\cite{ref-27,ref-28}—suffices for the present discussion. With this, we may calculate the changes in the saturation magnetization $M_s$ at finite temperature under different values of $<\epsilon_m>$ , controlled by the parameters $a$, $c$, and $\psi$. Such estimations are shown in \textbf{Figures 2(a)–2(c)} for the parameters $a$, $c$, and $\psi$, respectively. The exponential suppression of the low-energy density of states, resulting in an effective suppression of the magnetization loss, is clearly visible in these figures. The magnitude of the suppression increases as $<\epsilon_m>$ increases.
In various contexts, it has been shown that in the presence of disorder, the saturation magnetization often follows an expression slightly different from the usual Bloch equation and can be written as\cite{ref-23,ref-25,ref-29}:
\begin{equation}
M_s = M_0\left[ 1 - \frac{M_{loss}}{M_0}\right] =  M_0\left[1- C^\prime T^{3/2} \exp\left\{-BT^\beta\right\}\right]
\end{equation}
Where $B$ and $C^\prime$ are arbitrary constants, and $M_0$ is the saturation magnetization at zero temperature. While it is difficult to obtain such an expression analytically within the present model, we have attempted to simulate the saturation magnetization using an expression of the form:
\begin{equation}
\resizebox{.97\hsize}{!}
{$M_s = M_0\left[ 1 - \frac{M_{loss}}{M_0}\right] =  M_0\left[1- C T^{\alpha} \exp\left\{-BT^\beta\right\}F(H)\right]$}
\end{equation}
Where $F(H)$ arises from the magnetic-field dependence of $M_{loss}$, and for very low values of $H$ used in the calculations, it may be approximated as $\sim 1$. \textbf{Figure 2(d)} shows one such representative fit of $\ln(M_{loss})$ using the above expression with $F(H)=1$. In the same way, $M_{loss}$ obtained for different values of $<\epsilon_m>$, by varying the parameters $a$, $c$, and $\psi$, was fitted using the above expression. \textbf{Figure 3(a)} shows the values of $\alpha$ and $\beta$ obtained from such fittings. It is interesting to note that, within the uncertainties, the values of $\alpha$ and $\beta$ can be approximately taken as 0.5 and -0.5, respectively.
In order to determine the expression for $F(H)$, we calculate $M_{loss}$ as a function of $H$ at different temperatures for specific values of the parameters $a$, $c$, and $\psi$. Note that, to observe a substantial change in the presence of a magnetic field, one must choose a low value of $<\epsilon_m>$. The variation of $M_s$ with $H$ at a fixed temperature of approximately 50 K is shown in \textbf{Figure 3(b)}. Interestingly, we find that the variation of $M_{loss}$ with $H$ at any temperature between 5 K and 100 K can be fitted with an expression of the form:
\begin{equation}
M_{loss}(H) = D  \exp\left\{-\mu H/k_B T\right\}
\end{equation}
Here, $D$ is an arbitrary constant and $\mu$ denotes the magnetic moment.. One such representative fit is shown in \textbf{Figure 3(c)}. By combining this behavior in the presence of weak disorder with the above expression, one may write the overall expression for $M_s$ for a spin 1/2 system of any dimension in the weak-disorder limit as:
\begin{equation}
M_s =  M_0\left[1- A T^{0.5} \exp \left\{-B T^{-0.5}- \left(\mu H/k_B T\right)\right\}\right]
\end{equation}
Where $A$ and $B$ are arbitrary constants.
One may further extend this analogy to spin-transport phenomena\cite{ref-30,ref-31} in weakly disordered ferromagnetic systems. If one assumes that, in the weak-disorder limit, magnons can diffuse\cite{ref-31} or tunnel\cite{ref-32} across the boundaries created by these disorders, and that scattering due to these disorders, as well as magnon–magnon scattering, is reduced, then the spin current along a given direction due to magnon excitation may be written using the following simplified expression\cite{ref-30}:
\begin{equation}
j \propto \sum_k {v_k <n_k>}
\end{equation}
Where $v_k$ is the group velocity of the magnon for a particular mode $k$, and $<n_k>$ is the population. Thus, one may follow the same derivation presented earlier and show that, if it can be approximated that, under a discretized energy spectrum, $v$ is independent of $<n_\epsilon>$, then one may derive a similar expression for the spin current in the presence of weak disorder and in the weak-magnetic-field regime as:
\begin{equation}
j = A^\prime T^{0.5} \exp \left\{-B^\prime T^{-0.5}\right\}
\end{equation}
Where $A^\prime$ and $B^\prime$ are arbitrary constants. Note that this expression, though derived in a very different way, is structurally very similar to the Efros–Shklovskii equation\cite{ref-24}, which describes electron conductivity in the presence of disorder. Indeed, an increase in the spin current at low temperatures, following a $\sim T^{0.5}$ dependence and possible subsequent saturation, has been observed in selected ferromagnetic samples\cite{ref-33}.

While the above-mentioned derivation does reproduce the experimentally observed $\sim T^{0.5}$ dependence of the spin current, a few points should be noted here. First, in the above derivation, the mean of the distribution is unrealistically large in some cases. Note that the specific value of $<\epsilon_m>$ does not represent the lowest gap of the system; rather, it is a combined effect of the lowest gap, the variation of arbitrary $L$, and the magnon bandwidth driven by the $n^2$ term. One may impose restrictions on these parameters and obtain a mean value much lower than that presented here. However, such choices do not alter the main physics discussed in this work.\\
The second point to note is the use of the approximate Maxwell–Boltzmann approximation in the derivation. In this regard, note that various contrasting concepts have been proposed for magnon populations: in some cases, the Bose–Einstein (BE) distribution\cite{ref-26} is used, while in others a hard-core boson (HCB) model\cite{ref-34,ref-35,ref-36} with a Fermi–Dirac distribution has been employed. The use of the Maxwell–Boltzmann approximation provides an approximate way to reconcile such variations. However, in order to understand the effect of using different population models, we have performed the entire calculation using both the BE model and the HCB model while keeping the $<\epsilon_m>$ of the distribution lower and more realistic. Such calculations are shown in \textbf{Figures 4(a) and 4(b)}. Note that, even within this model, we may obtain approximate expressions for the saturation magnetization loss and spin current (in the low field limit) as follows:
\begin{equation}
M_s =  M_0\left[1- A T^{\alpha} \exp \left\{-B T^{\beta}- \left(C \mu H/k_B T\right)\right\}\right]
\end{equation}
\begin{equation}
j = A^\prime T^{\alpha} \exp \left\{-B^\prime T^{\beta}\right\}
\end{equation}
\begin{figure}[t]
\begin{center}
\includegraphics[width=1.0\columnwidth]{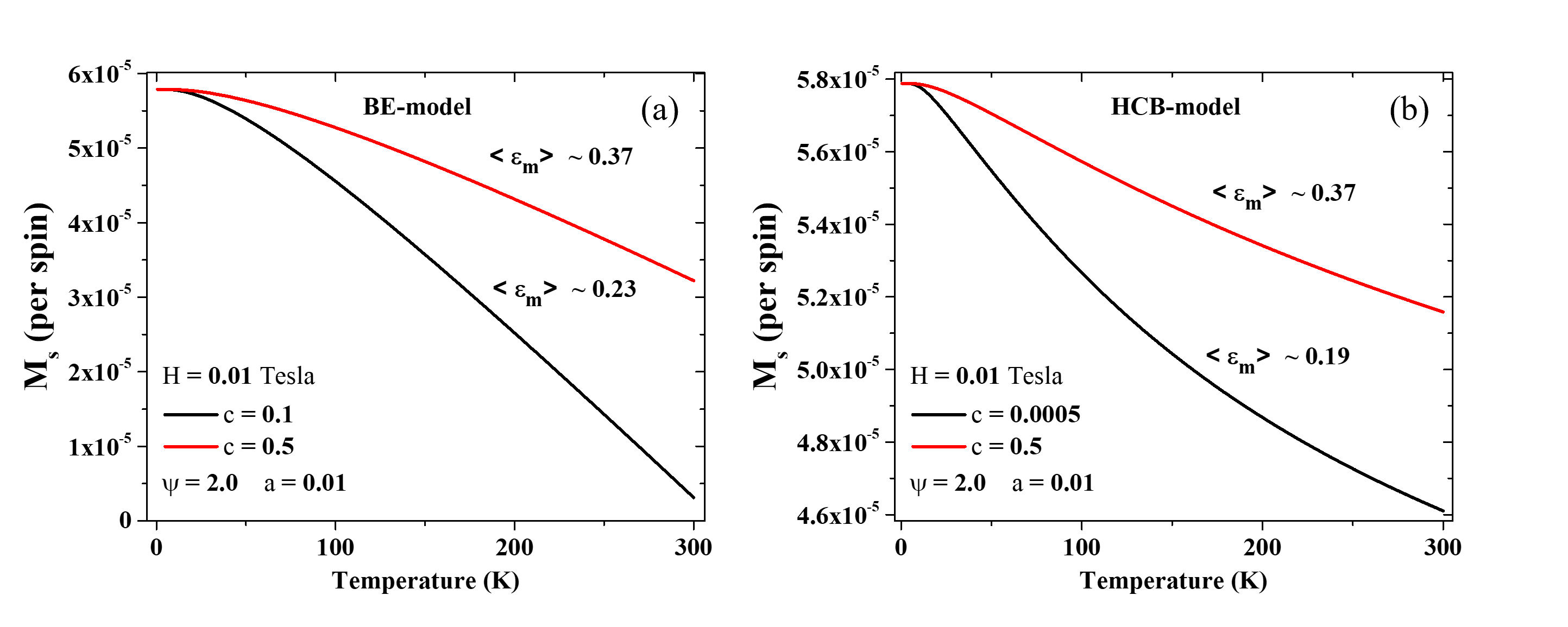}
\caption{(a)–(b) Calculated variations of the saturation magnetization ($M_s$) for two different sets of parameter values of $a$, $c$, and $\psi$, obtained using the BE and HCB models, respectively, at different temperatures under a fixed low magnetic field. Note that we do not extend this calculation to very low values of $<\epsilon_m>$, since the Mermin–Wagner effect becomes increasingly pronounced as the gap decreases.}
\label{fig4}
\end{center}
\end{figure}
Where $A$, $B$, $C$, $A^\prime$, $B^\prime$ are arbitrary constants, whereas $\alpha$ typically takes values between $\sim 0.5$ and $\sim 1.5$, and $\beta$ typically takes values between $\sim (- 0.5)$ and $\sim (- 1.0)$. However, as stated in the introductory section, these derivations are valid only in the low-temperature regime and are more applicable to systems where the phase transition is smeared by disorder. \\

In conclusion, we have adopted a simplified approach and derived a modified Bloch equation to describe the loss of saturation magnetization in a weakly disordered spin-1/2 ferromagnetic system, which may be applied to systems of varying dimensionality. This approach further allows us to obtain an approximate expression for the spin current in such systems.

\section{Acknoledgement}

The author thanks the Indian Institute of Science, Bangalore, for providing the facilities to conduct this work. Parts of the manuscript was proofread for grammatical accuracy using ChatGPT, an AI-based language tool.

\section{Data availability}

The data that support the findings of this study are available from the corresponding author upon reasonable request.



\end{document}